\documentclass{aa}  

\usepackage{graphicx}
\usepackage{float}
\usepackage{txfonts}
\usepackage[usenames,dvipsnames]{color}
\usepackage{keyval}
\usepackage{hyperref}

\usepackage{orcidlink}
\begin{document}

   \title{The attempted polarity reversal and evolving magnetic environment of AD Leo}

   \subtitle{}

   \author{K. G. Smith\inst{1,2}\orcidlink{0009-0002-4483-3701}
    \and
    D. Evensberget\inst{1,3}\orcidlink{0000-0001-7810-8028}
    \and
    S. Bellotti\inst{1,4}\orcidlink{0000-0002-2558-6920}
    \and
    J. Morin\inst{5}\orcidlink{0000-0002-4996-6901}
    \and 
    A. A. Vidotto\inst{1}\orcidlink{0000-0001-5371-2675}
    \and 
    B. J. S. Pope\inst{6,2}\orcidlink{0000-0003-2595-9114}
    }

   \institute{
   Leiden Observatory, Leiden University, PO Box 9513, 2300 RA Leiden, The Netherlands\\
    \email{bellotti@strw.leidenuniv.nl}
    \and 
   School of Mathematics and Physics, University of Queensland, St Lucia, QLD 4072, Australia
    \and
    Centre for Planetary Habitability (PHAB), Department for Geosciences, University of Oslo, Oslo, Norway
    \and
    Institut de Recherche en Astrophysique et Plan\'{e}tologie, Universit\'{e} de Toulouse, CNRS, IRAP/UMR 5277, 14 avenue Edouard Belin, F-31400, Toulouse, France
    \and 
    Laboratoire Univers et Particules de Montpellier, Universit\'{e} de Montpellier, CNRS, F-34095, Montpellier, France
    \and 
    School of Mathematical and Physical Sciences, Macquarie University, Macquarie Park, NSW 2113, Australia}

   \date{Received XXX; accepted XXX}

% \abstract{}{}{}{}{} 
% 5 {} token are mandatory
  \abstract
  {In the past two decades, the observed large-scale magnetic field of the active M dwarf star AD~Leo has evolved from being strongly negative to mildly negative, raising a suspicion that it might be at the imminence of switching polarity (i.e. becoming positive). Although magnetic field reversals are observed every 11 years in the solar magnetic field, in the context of M dwarfs, magnetic field reversals are still poorly understood and so far not predictable. Further, no reversals have yet been observed for fast-rotating M dwarfs. Moreover, it is known that the magnetic field of stars impacts their surrounding space weather environment. Studying how space weather evolves over time is thus crucial for examining planetary habitability.}
  {We examine the properties of AD~Leo's large-scale magnetic field, which was recently found to be trending towards a polarity reversal. We also investigate how the space weather environment changes in response to the evolution of the large-scale magnetic field, by modelling the wind of AD~Leo.}
  {We analysed spectropolarimetric data collected by ESPaDOnS and SPIRou in late 2022 and early 2023. With the optical and near-infrared data we computed the longitudinal magnetic field, and with the near-infrared data we reconstructed the large-scale magnetic field using Zeeman-Doppler imaging. Using five magnetograms, from between 2019 and 2023, we simulated three-dimensional Alfvén wave-driven stellar winds using the state-of-the-art space weather code \texttt{SWMF}.}
  {Although we see an evolution of the large-scale magnetic field of AD~Leo, we find no polarity reversal, but rather a restoration of the field to a simpler and axisymmetric configuration and consistently negative values for the longitudinal magnetic field strength. Previous work found the longitudinal field to get as weak as $-46$\,G in 2020, and rather than continued weakening it now appears to be strengthening. Our new large-scale field reconstruction for AD~Leo is characterised by a highly axisymmetric, poloidal-dipolar field with an increased mean large-scale field strength from 93\,G to 145\,G. However, the mean strength is still diminished compared to maps produced between 2007 and 2016. Our simulations of the space weather environment around AD~Leo find the stellar mass loss rates to be on average $\sim2.54\times10^{10}$\,kg/s -- an order of magnitude greater than the solar mass loss rate. Additionally, we examine the space weather experienced by hypothetical planets orbiting at the bounds of the habitable zone around AD~Leo. We find that the entirety of the habitable zone resides beyond the Alfvén surface. Further, magnetised habitable zone planets (with planetary field strengths greater than 0.34\,G) would likely be shielded from the incident wind and atmospheric erosion would be negligible (excluding effects from coronal mass ejections and flares). Additionally, we find the complexity of the wind velocity and wind pressure structures to evolve with the changing axisymmetry of the stellar large-scale magnetic field, resulting in more variable conditions along the orbits at certain epochs.}
  {}


   \maketitle

%
%-------------------------------------------------------------------

\section{Introduction}

M dwarfs constitute the vast majority of stars in the solar neighbourhood. They are expected to host an abundance of temperate and terrestrial planets~\citep{Ment2023}, making them prime candidates for habitability studies. The lifetimes of these stars exceed those of solar-like stars by at least a factor of 20, as they burn through their nuclear fuel at a slower rate~\citep{Adams1997}. M dwarfs experience higher levels of magnetic activity, relative to solar-like stars, which can establish hostile space weather environments that may compromise planetary habitability~\citep{Tilley2019}. The relatively low masses (0.08\,M$_\odot$ to 0.62\,M$_\odot$) of these stars result in correspondingly low luminosities ($2.4\times10^{-4}$\,L$_\odot$ to $7.6\times10^{-2}$\,L$_\odot$) and effective temperatures (2,300\,K to 3,900\,K), and thus relatively close-in circumstellar habitable zones~\citep{Kasting1993,Shields2016}. The close orbits of planets about these stars increase the risks posed by stellar activity such as flares and coronal mass ejections (CMEs).

Stellar activity is the outward manifestation of the internal dynamo. The nearly periodic variation in the Sun's activity is described by the 11 year Schwabe cycle~\citep{Schwabe1844, Hathaway2015}. Throughout the solar cycle, we observe changes in activity, such as the rates of flaring, CMEs, and the occurrence of sunspots~\citep{Gnevyshev1966}. Specifically, we observe variations in the occurrence rates of sunspots, accompanied by latitudinal migrations~\citep{Karoff2019}. With every 11 year cycle, we also note that a polarity reversal of the Sun's large-scale field is facilitated by a reversal of the leading spot polarity in each hemisphere~\citep{Hathaway2015}. Therefore, we also have a 22 year magnetic cycle, known as the Hale cycle~\citep{Hale1925}. This magnetic cycle is also reflected in variations in the large-scale field's tilt angle, as well as the contributions of the poloidal and toroidal field components to the overall strength of the field~\citep{Parker1955, Vidotto2018}. Similar activity cycles have been identified for several solar-like stars~\citep[see e.g.][]{Sanderson2003, Jeffers2023, Bellotti2025}.

The magnetic dynamo of the Sun is described by an $\alpha\Omega$ effect (of the Babcock-Leighton type; \citealt{Cameron2016}. From this theory, cyclical variations in the Sun's magnetic field arise from the interaction between differential rotation ($\Omega$ effect) and the helical motion, by means of the Coriolis force, of plasma ($\alpha$ effect) in the convection zone~\citep{Bao2002}. Solar and stellar dynamo is an active field of research, as a consensus on the exact underlying mechanism has not yet been reached. It is widely thought, however, that the Sun's dynamo is related to its internal structure. With this, M dwarfs are interesting laboratories for testing the robustness of dynamo theories, since we can separate them into partially convective ($>$0.35\,$M_\odot$) and fully convective ($<$0.35\,$M_\odot$) populations~\citep{Baraffe2018}. Partially convective M dwarfs exhibit a similar internal structure to the Sun, consisting of a radiative core and a convective envelope. These layers are separated by the tachocline, which is believed to be crucial to the dynamo processes of our Sun~\citep{Thompson2003, Duguid2023}. Fully convective M dwarfs, however, do not possess a radiative core, and thus contain no tachocline.

Detecting and reconstructing large-scale stellar magnetic fields relies on spectropolarimetry. In particular, one can use Zeeman-Doppler imaging (ZDI) to map the large-scale photospheric field~\citep{Donati1997b} from a time series of high-resolution spectropolarimetric observations. Thus far, observational studies of M dwarfs have revealed potential polarity reversals for two slow-rotating stars, GJ 1151 (M = 0.17 \textpm ~0.02\,$M_\odot$, $P_\text{rot}$ = 140 \textpm~10\,d) and Gl 905~\citep[M = 0.15 \textpm ~0.02\,$M_\odot$, $P_\text{rot}$ = 114.3 \textpm~2.8\,d;][]{Cristofari2022, Donati2023b, BlancoPozo2023, Lehmann2024}. Although such reversals have not yet been observed for fast-rotating M dwarfs, recent work has hinted at an imminent polarity reversal for AD~Leonis~\citep[M = 0.42\,$M_\odot$, $P_\text{rot}$ = 2.2399 \textpm 0.0006\,d;][]{Morin2008, Mann2015,Cristofari2023} -- alluded to by significant reductions in the axisymmetry of the large-scale field~\citep{Bellotti2023b}. Previous studies also suggested a global weakening in field strength over time~\citep{Lavail2018}. It should be noted that, while reconstructions of the large-scale field made using ZDI suggest this weakening, the same trend has not been observed with surface magnetic field measurements based on Zeeman broadening modelling \citep[with the surface field almost consistently exhibiting strengths greater than 3\,kG;][]{Bellotti2023b}.

Crucially, for our Sun, the activity cycle modulates the space weather conditions in the Solar System. The magnetic solar dynamo results in fluctuations in the solar wind velocity, and thus we observe space-climatic feedback in response to the solar cycle~\citep{Carbone2024}. Further, the spin-down process of the Sun is enhanced by the magnetised solar wind, this comprising streams of magnetised particles emanating from the upper atmosphere of the Sun~\citep{Schatzman1962, Chebly2023, Evensberget2024}. These winds are generally characterised by a steady flow that is occasionally punctuated by transient, high-energy events \citep{Gombosi2018, Mishra2019}. In general, solar wind can be categorised as fast (with a median speed of 760\,km\,s$^{-1}$) or slow \citep[with a median speed of 400\,km\,s$^{-1}$;][]{McComas2003, Johnstone2015}. The two components are linked to the large-scale magnetic field geometry. Given that the magnetic field of the Sun evolves throughout the solar cycle, so too do the origins of these fast and slow winds~\citep{Krieger1973,Saikia2020}, which correlate with changes in angular momentum loss rates throughout the solar cycle~\citep{Cohen2011,Finley2018, Evensberget2021, Evensberget2024}. Not only do we observe changes in the quiescent winds of the Sun, but also the severity and frequency of transient solar activity events, such as flares and CMEs, are modulated by the solar cycle~\citep{Chapman2020, Owens2021}.

The space weather in the traditional habitable zone around M dwarfs may be too harsh to sustain life~\citep{Tilley2019}. The presence of a planetary magnetic field can shield the planetary atmosphere from direct interaction with stellar winds and so may play a crucial role in atmospheric retention~\cite{Stern1982, Moore2007, RodriguezMozos2019}. Understanding the broader implications of space weather on exoplanets is essential for understanding habitability. Variability in stellar winds, flares, and CMEs can have drastic impacts on planetary systems. In the Solar System, for example, interactions between Venus' ionosphere and solar wind result in an induced magnetosphere that envelops the planet~\citep{Luhmann2004}. For CO$_2$-rich rocky exoplanets, in the habitable zone of low-mass M dwarfs, up to tens of bars of atmospheric gas may be lost as a result of stellar plasma interactions over time~\citep{Lammer2007}. In the case of these planets, a Venus-like induced magnetosphere may result in enhanced planetary atmospheric escape rates~\citep{Zhang2009}. Thus, studying the space weather environment over time is crucial for understanding how exoplanetary atmospheres may be eroded or sustained over geological timescales~\citep{Airapetian2020}. The space weather modelling intended to characterise the severity of such environments requires a sound understanding of the large-scale magnetic field topology of the star \citep{Vidotto2023}. 

AD~Leo is a highly active M dwarf that has been the subject of a number of flare studies. In this work, we turn our attention to quiescent space weather. We used new spectropolarimetric observations of AD~Leo to characterise the large-scale field, using ZDI, and re-examined the plausibility of a polarity reversal. Using five magnetograms, we simulated the quiescent space weather environment around this star to understand how it evolves across epochs in response to a changing large-scale magnetic field. While not the primary focus of this work, we also consider the potential impacts of the simulated space weather on hypothetical planets orbiting in the habitable zone. We note that AD~Leo is an extreme case for considering habitability and, as such, is not representative of a typical exoplanet host star.

This paper is structured as follows. In Sect.~\ref{sec: Obs} we detail the observations of AD~Leo in both near-infrared and optical domains. In Sect.~\ref{sec: large-scale field} we detail our analysis of AD~Leo’s large-scale field, including calculations of the longitudinal magnetic field and a topological reconstruction. In Sect.~\ref{sec: winds} we discuss the wind modelling results, and in Sect.~\ref{sec: sum} we present our conclusions.

%--------------------------------------------------------------------
\section{Observations \label{sec: Obs}}
AD~Leo (M3.5) is one of the closest and brightest M dwarfs, at a distance of $4.9650\pm0.0007$\,pc~\citep{Gaia2020} and with V and H band magnitudes of 9.52 and 4.84, respectively~\citep{Cutri2003,Zacharias2012}. It is an active star that has been studied extensively in both its quiescent and flaring states \citep[e.g.][]{Pettersen1984,SanzForcada2002,Maggio2004, Mohan2024}. With its rotation period of 2.2399\,days~\citep{Morin2008, Fouque2018} and quiescent X-ray luminosity of $L_\text{x}/L_\text{bol}=9\times10^{-4}$~\citep{Giampapa1996} it is situated within the saturated regime of the rotation-activity relation~\citep{Robrade2005, Wright2011, Stelzer2022}. The mass of AD~Leo (0.42\,$M_\odot$) places it in the partially convective regime~\citep{Mann2015}.

In later sections, the following expression will be used to phase the observations:
\begin{align}
    HJD = HJD_{\text{ref}} + P_{\text{rot}} \cdot n_{\text{cyc}}, \label{eq: ephemeris}
\end{align}
where we take $P_{\text{rot}}$ to be the stellar rotation period of AD~Leo (2.2399\,d) and $n_{\text{cyc}}$ to be the rotation cycle of the observation. The reference heliocentric Julian date for the new observations is $HJD_{\text{ref}} = 2459894.15$.

\subsection{Near-infrared data}
Between November 2022 and January 2023, 19 spectropolarimetric observations of AD~Leo were made using the \textit{SpectroPolarimètre InfraRouge} (SPIRou). SPIRou is a near-infrared spectropolarimeter installed on the 3.6 m Canada-France-Hawaii Telescope (CFHT). This instrument covers a wavelength domain from 0.967\,\textmu{}m to 2.493\,\textmu m with a resolving power of 70,000. The observed spectra were processed with A PipelinE to Reduce Observations (APERO v0.7.288), which delivers telluric-corrected spectra~\citep{Cook2022}. 

Observations were made in the circular polarisation mode and spanned a period of 64 days, with their record provided in Table~\ref{tab: obs}. The mean airmass was 1.10 and the signal-to-noise ratio (S/N), per spectral resolution element, at a wavelength of 1650 nm ranged from 261 to 477, with an average of 388. We assume the observed spectrum to be the convolution between a line mask and the mean line profile, and so we used the least-squares deconvolution \citep[LSD;][]{Kochukhov2010, Donati1997} technique to obtain the mean Stokes $I$ (unpolarised) and Stokes $V$ (circularly polarised) profiles. We adopted a velocity step of 1.8~km/s for the computation of the LSD profiles and used the \texttt{LSDpy} code that is part of the \texttt{specpolflow} package \citep{Folsom2025}. 

The line mask describes a comb of weighted Dirac delta functions centred at absorption lines in the spectrum and encoding information about properties such as depths and Landé factors (that is, a quantification of the magnetic sensitivity of a line). The mean line profiles produced from the LSD process combined information from many spectral lines observed and enhanced the S/N, allowing polarimetric signatures to be extracted. The line mask implemented was produced with the Vienna Atomic Line Database~\citep[VALD,][]{Ryabchikova2015} and a MARCS atmosphere model~\citep{Gustafsson2008}, with parameters $T_{\text{eff}}=3500$\,K, $\log{g}=5.0$\,cm\,s$^{-2}$, and $v_{\text{micro}}=1.0$\,km\,s$^{-1}$. It was characterised by 1399 atomic lines of wavelengths ranging from 950 nm to 2600 nm, with known Landé factors (roughly ranging from 0 to 3) and depths greater than 3\% of the continuum level. 

For our analysis, we used 1242 of the lines from the mask that were within the wavelength coverage of our observations. The telluric bands affecting NIR observations were as follows: 950\,nm to 979\,nm, 1116\,nm to 1163\,nm, 1331\,nm to 1490\,nm, 1790\,nm to 1985\,nm, 1995\,nm to 2029\,nm, and 2250\,nm to 2500\,nm. However, the effects of these bands were handled by the robust telluric corrections put in place by SPIRou such that they should not affect data analysis~\citep{Carmona2023, Bellotti2023b}. As a precaution, we opt to remove the telluric bands, prior to computing the LSD profiles, to account for any potential residuals following the telluric correction.

\begin{figure}[t]
    \centering
    \includegraphics[width=1.0\linewidth]{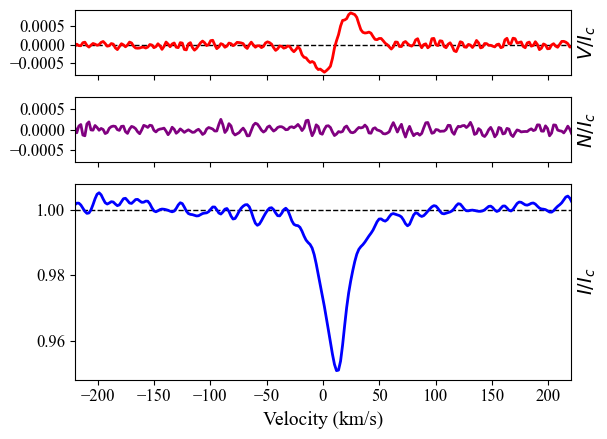}
    \caption{Normalised Stokes LSD profiles from a SPIRou observation of AD~Leo on 13 January 2023. The top panel displays the Stokes $V$ (circularly polarised) profile; the middle panel displays the Null profile (a measurement of spurious polarisation signals); the bottom panel displays the Stokes $I$ (unpolarised) profile.}
    \label{fig: LSD_profs}
\end{figure}

Figure~\ref{fig: LSD_profs} shows the results of LSD applied to the observation on 13 January 2023. We observe a two-lobed antisymmetric structure that is characteristic of observing the negative polarity of a dipolar large-scale topology~\citep[e.g.][]{Bellotti2023b}. The average noise, relative to the unpolarised continuum, for the Stokes $V$ profiles from the near-infrared observations is $8.97\times10^{-5}$.  

\subsection{Optical data}
In October 2022, over a 4 day period, four optical spectropolarimetric observations of AD~Leo were made with the \textit{Echelle SpectroPolarimetric Device for the Observation of Stars} (ESPaDOnS). This instrument provided coverage of the optical spectrum from 370\,nm to 1050\,nm, with a resolving power of 68,000. A journal of these observations is provided in Table~\ref{tab: obs}. Much like SPIRou, ESPaDOnS is mounted on the CFHT. The data reduction pipeline for this instrument was Libre-ESpRIT (v2.12; \citealt{Donati1997}). 

As with the near-infrared data, we retrieved the line mask from VALD and computed the Stokes LSD profiles (once again with a velocity step of 1.8~km/s). The optical mask contained 4216 atomic lines in the approximate wavelength range of 350 nm $-$ 1080\,nm, with depths greater than 40\% the continuum level. For our analysis, we used 3300 atomic lines. The difference between the number of atomic lines in the mask and the number of lines implemented arose from the removal of regions of the spectrum that were either impacted by telluric lines or in the vicinity of H$\alpha$. These regions were: 627\,nm to 632\,nm, 655.5\,nm to  657\,nm, 686\,nm to 697\,nm, 716\,nm to 734\,nm, 759\,nm to 770\,nm, 813\,nm to 835\,nm, and 895\,nm to 986\,nm, following \citet{Bellotti2022}. For the optical observations used in this study, the average noise of the Stokes $V$ profiles was $1.6\times10^{-4}$ relative to the unpolarised spectrum.

\section{Large-scale magnetic field \label{sec: large-scale field}}
\subsection{Longitudinal magnetic field}
For each of the observations made, the line-of-sight component of the disc-integrated magnetic field ($B_\ell$) could be calculated from the LSD Stokes $I$ and $V$ profiles. Measurements of the mean longitudinal magnetic field have been used as tracers for stellar activity. This quantity is computed as
\begin{align}
    B_{\ell} = - \frac{h}{\mu_B\langle \lambda \rangle \langle g \rangle}\frac{\int vV(v) ~\mathrm{d}v}{\int (I_c - I)~\mathrm{d}v}, \label{eq: Bl}
\end{align}
where $h$ is Planck's constant, $\mu_B$ is the Bohr magneton, $\langle \lambda \rangle$ and $\langle g \rangle$ represent the normalisation wavelength and normalisation Landé factor of the LSD profiles, $I_c$ is the continuum level, and $v$ denotes the radial velocity in the stellar reference frame. It can also be noted that $hc/\mu_B=0.0214$\,Tm, with $c$ being the speed of light in a vacuum. In the case of the near-infrared observations, the normalisation wavelength and Landé factor had values of 1700\,nm and 1.24, respectively. For the optical observations, their values were 700 nm and 1.25. Further, the integration ranges for the near-infrared and optical LSD profiles were $\pm50$\,km\,s$^{-1}$ and $\pm30$\,km\,s$^{-1}$, respectively, relative to the centre of the Stokes \textit{I} profiles at 11.71\,km/s. The strength of the Zeeman effect (and thus the separation between the lobes of the Stokes \textit{V} profile) scales as $\lambda^2$, and so a wider integration range is required in the near-infrared to capture the entirety of the signal.

\begin{figure}[t]
    \centering
    \includegraphics[width=1\linewidth]{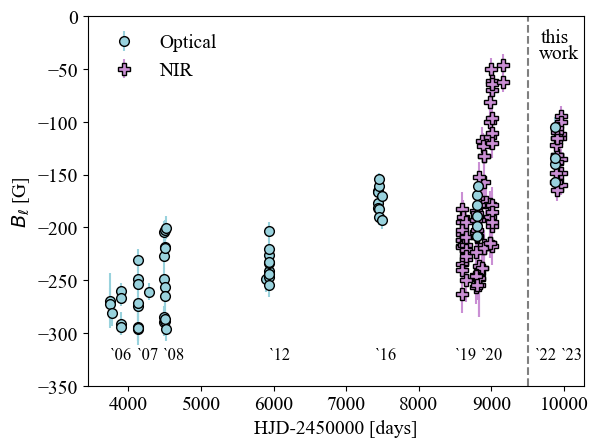}
    \caption{Temporal evolution of the longitudinal magnetic field ($B_\ell$) from 2006 to 2023, using observations from ESPaDOnS, Narval, and SPIRou. $B_\ell$ and HJD measurements prior to 2023 were sourced from \citet{Morin2008}, \citet{Lavail2018}, and \citet{Bellotti2023b}. The blue circles represent observations made in the optical domain, meanwhile the pink pluses represent near-infrared observations. The text below the data points roughly indicates the year of the observations.}
    \label{fig: Bl}
\end{figure}

As is shown in Fig.~\ref{fig: Bl}, we report consistently negative values for the longitudinal magnetic field, suggesting we are observing a single pole of a dipole whose magnetic axis is largely aligned with the axis of rotation; given that AD~Leo's inclination is only 20 degrees. The values for the near-infrared measurements range from $-164$\,G to $-94$\,G, with an average of $-128$\,G and median error of $10$\,G. Meanwhile, the four optical measurements possess values of $-157$\,G to $-105$\,G, with an average of $-134$\,G and a median error of $9$\,G.

Following the work of \citet{Bellotti2023b}, the $B_\ell$ results for these more recent spectropolarimetric observations subvert expectations regarding the long-term evolution in the longitudinal field strength. This previous work found the field strength to get as weak as $-46$\,G, from near-infrared observations made in 2020. Rather than continued weakening of the longitudinal field, we find it appears to be strengthening again. Phase-folding these new $B_\ell$ measurements, in accordance with Eq.~\ref{eq: ephemeris}, suggests little evidence for rotational modulation (as shown in Fig.~\ref{fig: phasefolded}) -- in agreement with an increase in axisymmetry since previous observations.

\subsection{Topological reconstruction \label{sec: reconstruction}}
We expect little variability in the large-scale field configuration of AD~Leo on the temporal scale of a few months~\citep{Bellotti2023b}. As such, we are able to produce a single large-scale field reconstruction informed by observations from November 2022 and January 2023.

With the time series of circularly polarised Stokes $V$ profiles from the near-infrared SPIRou observations, we recovered the large-scale magnetic field topology at the stellar surface using ZDI~\citep{Donati1997b}. We note there were too few ESPaDOnS observations to produce a new surface magnetogram in this wavelength domain. The resulting geometry is described in terms of spherical harmonics and, due to the divergence free nature of the large-scale field, is expressed as the sum of a poloidal and a toroidal component. The implementation of ZDI in \texttt{ZDIpy}\footnote{The \texttt{ZDIpy} code is available at: \url{https://github.com/folsomcp/ZDIpy}} applies a maximum entropy regularisation that allows us to obtain a large-scale magnetic field map describing the data and with minimal information content~\citep[for more details see][]{Skilling1984, Donati1997b, Folsom2018}. The final large-scale magnetic field models were obtained by iteratively comparing observed and synthetic Stokes $V$ profiles until the target reduced $\chi^2$ was reached. The value of the target reduced $\chi^2$ was selected such that it prevented over-fitting. The method implemented for its selection is described in \citet{AlvaradoGomez2015}.   

For this work, we applied the Unno-Rachkovsky's solutions to the equations for polarised radiative transfer in a Milne-Eddington atmosphere~\citep{Unno1956, Rachkovsky1967,LandiDegl'Innocenti2004, Bellotti2023b} as well as the filling factor formalism detailed in \citet{Morin2008}. This formalism allows each surface element to contain both magnetic and non-magnetic regions. This way, we assume that small-scale magnetic spots, arranged according to the large-scale magnetic field configuration, generate the polarisation signature. This formalism allows us to account for the fact that ZDI is insensitive to small-scale magnetic structures (i.e. the primary contributors to the global magnetic field strength of the star). The local magnetic field of these spots is $B/f_V$ and the magnetic field modulus averaged over the cell is the ZDI-reconstructed $B$. We performed an optimisation of the Stokes~$V$ filling factor following \citet{Bellotti2023b} but it was inconclusive; hence, we set it to a value of 10\%, which is consistent with the range (9\%--16\%) found by \citet{Bellotti2023b}. In Sect.~\ref{sec: boundary_conditions} we discuss the impact of our choice for the Stokes~\textit{V} filling factor on the wind modelling results. 

In our model, we assumed $i=20^\circ$, $v_e\sin{(i)}=3$\,km\,s$^{-1}$, $P_{\text{rot}}=2.2399$\,d, and solid body rotation \citep[following][]{Morin2008, Bellotti2023b}, with a linear limb-darkening coefficient of $\mu=0.2$ in the H band~\citep{Claret2011}. Specifically, the Unno-Rachkovsky local line profiles are described by the Gaussian width ($w_G$), the Lorentzian width ($w_L$), line strength ($\eta_0$), and the slope of the source function in the Milne-Eddington atmosphere ($\beta$). The first three quantities were chosen by $\chi^2$ optimisation (as was done for the filling factor on Stokes $V$) and were found to be $w_G=2.5$\,km/s, $w_L\approx9.4$\,km/s, and $\eta_0\approx16.1$. The value of $\beta$ was derived from~\citep{Bellotti2024b, Erba2024}
\begin{align}
    \beta = \mu/(1-\mu).
\end{align}
Since we have $\mu=0.2$, we get $\beta=0.25$.
Lastly, the maximum degree of harmonic expansion was set to $\ell_{\text{max}}=5$. This limit is appropriate considering the low $v_e\sin(i)$ for this star.

\begin{figure}
    \centering
    \includegraphics[width=0.9\linewidth]{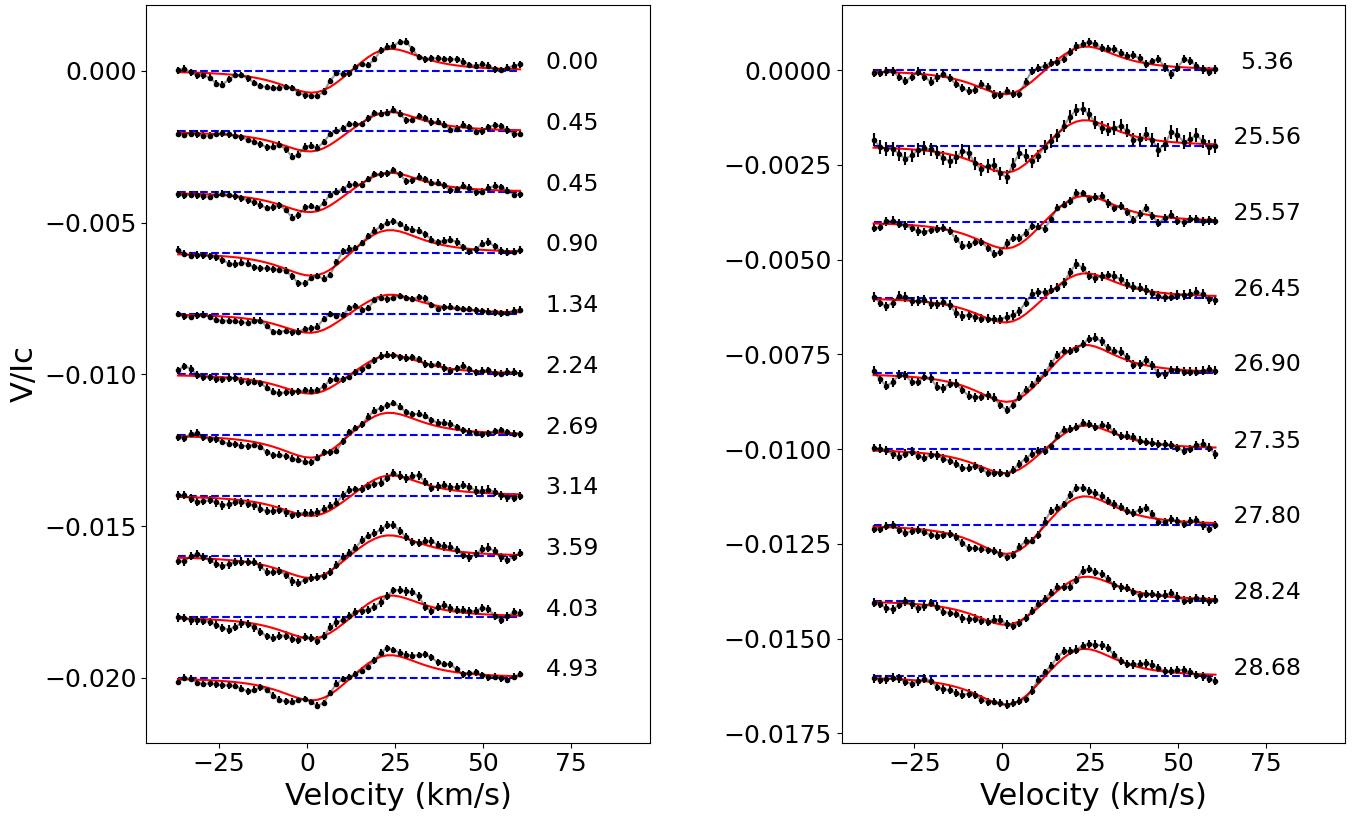}
    \caption{Full SPIRou 2022.9 time series of Stokes $V$ profiles for AD~Leo, normalised by the unpolarised continuum intensity. The observed profiles are presented in black, meanwhile the Unno-Rachkovsky fitted models are shown in red. The profiles are sorted with respect to their rotation cycle (see Eq.~\ref{eq: ephemeris}).}
    \label{fig: Stokes V fits}
\end{figure}

A time series of Unno-Rachkovsky models fit to the NIR circularly polarised profiles is displayed in Fig.~\ref{fig: Stokes V fits}. For these results, we fit the observations to a reduced $\chi^2$ of 1.18 (with an initial reduced $\chi^2$ of 13.52). The observed Stokes $V$ profiles display high-frequency variations about the models, though there are no significant deviations.

\begin{figure}
    \centering
    \includegraphics[width=1\linewidth]{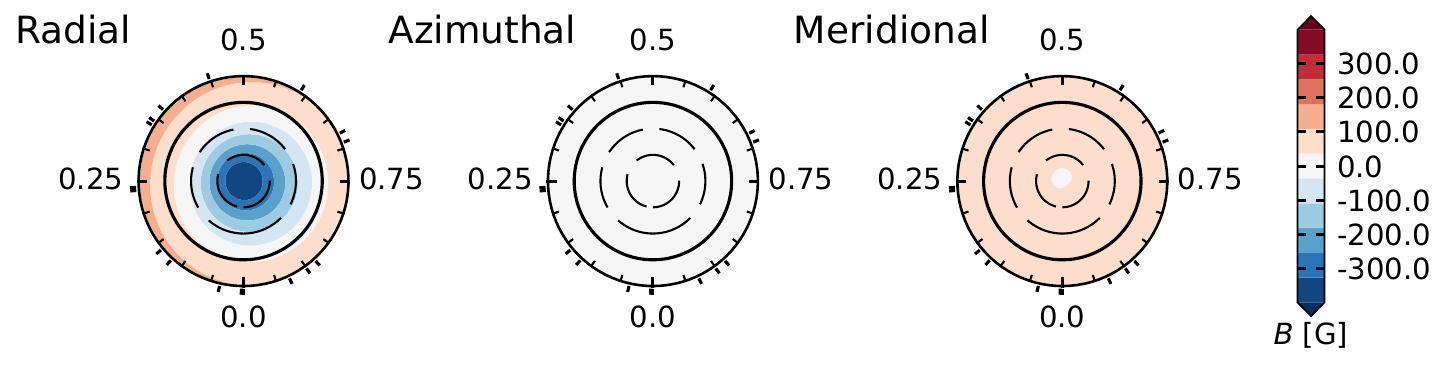}
    \caption{ZDI reconstruction of the large-scale magnetic field of AD~Leo from 2022.9 SPIRou observations, represented in a flattened polar view. We present the radial (left), azimuthal (middle), and meridional (right) components of the large-scale magnetic field vector. The radial ticks indicate the rotational phases at which observations were made, meanwhile the concentric circles indicate stellar latitudes: +60$^\circ$ (innermost), +30$^\circ$, equator, -30$^\circ$ (outermost). The range of the colour bar is determined by the maximum (in absolute value) strength of the large-scale magnetic field, with the negative and positive polarities represents by blue and red, respectively.}
    \label{fig: map}
\end{figure}

A map of the large-scale magnetic field is displayed in Fig.~\ref{fig: map} and its properties, along with the properties of previous large-scale field reconstructions, are provided in Table~\ref{tab: zdi_output}. This new reconstruction features a stronger meridional component than the maps produced in \citet{Bellotti2023b}. This discrepancy arises due to the use of a different spherical harmonics formalism. In this work, we employ the formalism of \citet{Lehmann2022}.

Our most recent configuration is predominantly poloidal (at 99.9\%), with the dipolar contribution accounting for 91\% of the poloidal magnetic energy - comparable to maps produced in 2012.0 and 2016.2~\citep{Lavail2018}. We also note an evolution in the contributions to the magnetic energy across the axisymmetric modes. There appears to be significant variation in the contributions of the dipolar and quadrupolar modes to the magnetic energy. The large-scale field was its least dipolar in 2007.1 and its most dipolar in 2016.2. Meanwhile, its quadrupolar component was weakest in 2016.2 but strongest in 2020.6. Further, we observe an evolution in the mean large-scale field strength and axisymmetry. The mean strength was greatest in 2012.0 (330.0\,G) and weakest in 2020.6 (93.4\,G). However, this weakening between 2012 and 2020 was not a steady decline. As for the axisymmetry, this quantity was its greatest in 2019.4 (99.8\%) and its lowest in 2020.6 (58.3\%). Since 2020.6, an epoch defined by a highly tilted and weakened large-scale field, AD~Leo's large-scale field has returned to a simpler, aligned configuration -- although still weak when considering the maximum reported large-scale field strength -- more reminiscent of the reconstructions produced from the 2019.4 observations. This work reveals a now highly axisymmetric large-scale field (97.9\%), a relatively low dipole obliquity (7.5\%) relative to the rotation axis, and a dipolar contribution of 90.9\%.

\begin{table*}[!t]
\caption{Properties of the magnetic maps of AD~Leo.} 
\label{tab: zdi_output}     
\centering                       
\begin{tabular}{l r r r r r r r r r r}      
\hline     
& 2007.1 & 2008.1 & 2012.0 & 2016.2 & 2019.4 & 2019.9 & 2019.9 & 2020.1 & 2020.6 & 2022.9\\ 
& VIS & VIS & VIS & VIS & NIR & NIR & VIS & NIR & NIR & NIR\\
\hline
$\langle B\rangle$ [G] & 190.0 & 180.0 & 330.0 & 300.0 & 111.2 & 132.3 & 158.0 & 115.3 & 93.4 & 144.8 \\
B$_\mathrm{max}$ [G]  & 1300.0 & 1300.0 & 1180.0 & 950.0 & 481.2 & 764.0 & 577.2 & 555.1 & 434.3 & 399.5\\
$\langle B^2\rangle$ [$\times10^{5}\,$G$^2$] & $0.61$ & $0.61$ & $1.20$ & $1.00$ & $0.19$ & $0.32$ & $0.36$ & $0.22$ & $0.14$ & $0.25$\\
f$_\mathrm{pol}$ [\%] & 99.0 & 95.0 & 94.0 & 90.6 & 100.0 & 99.9  & 95.0  & 99.3  & 98.7 & 99.9\\
f$_\mathrm{tor}$ [\%] & 1.0 & 5.0 & 6.0 & 9.4 & 0.0   & 0.1   & 5.0   & 0.7   & 1.3 & 0.1\\
f$_\mathrm{dip}$ [\%] & 56.0 & 63.0 & 89.0 & 94.1 & 81.3  & 71.1  & 81.7  & 75.7  & 70.1 & 90.9\\
f$_\mathrm{quad}$ [\%] & 12.0 & 9.0 & 4.4 & 2.0 & 14.9  & 19.0  & 14.6  & 17.7  & 21.2 & 7.4\\
f$_\mathrm{oct}$ [\%] & 5.0 & 3.0 & 1.3 & 1.1 & 2.8 & 6.2 & 2.7 & 4.4 & 5.9 & 1.1\\
f$_\mathrm{axisym}$ [\%] & 97.0 & 88.0 & 92.6 & 97.2 & 99.8  & 94.5  & 77.0  & 85.8  & 58.3 & 97.9\\
Obliquity [$^{\circ}$] & 8.0 & 17.5 &  13.5 & 8.0  & 2.5   & 12.5  & 21.5  & 19.5  & 38.0 & 7.5\\
\hline                                 
\end{tabular}
\tablefoot{The magnetic maps data are from 2007.1 and 2008.1~\citep{Morin2008}, 2012.0 and 2016.2~\citep{Lavail2018}, 2019.4, 2019.9 (NIR and VIS), 2020.1, and 2020.6~\citep{Bellotti2023b}, and 2022.9 (present work). The following quantities are listed: mean large-scale magnetic field strength (in absolute value), maximum large-scale magnetic field strength (in absolute value), the total magnetic energy, the contributions of the poloidal and toroidal magnetic energies to the total energy, the contributions of the dipolar, quadrupolar, and octupolar magnetic energies to the poloidal energy, the axisymmetric magnetic energy as a fraction of the total energy, and the tilt of the magnetic axis with respect to the rotation axis.}
\end{table*}

The ZDI maps presented here represent only the large-scale component of AD Leo's magnetic field. As will be discussed in Sect. \ref{sec: underestimation}, ZDI is insensitive to the small-scale structures of the magnetic field. For AD~Leo, Zeeman broadening measurements indicate a mean total surface field strength of approximately 3\,kG, which is substantially stronger than the sub-300\,G strengths recovered by ZDI~\citep{Bellotti2023b}. In the following sections, we focus on the large-scale magnetic field as the key component governing global wind topology and open magnetic flux at large-scale distances.

\section{ZDI-based wind modelling \label{sec: winds}}
In this section we present a time series of wind modelling results for AD~Leo. Using the Alfvén Wave Solar Model~\citep[\texttt{AWSoM},][]{Sokolov2013,vanderHolst2014}, an extension of \texttt{BATS-R-US}~\citep{Powell1999} within the Space Weather Modelling Framework~\citep[\texttt{SWMF},][]{Toth2005, Toth2012}, we simulated stellar winds using the magnetic maps obtained at epochs 2019.4, 2019.9, 2020.1, 2020.6, and 2022.9, from SPIRou observations. With \texttt{AWSoM}, a three-dimensional simulation was evolved forwards in time until we obtained a steady or quasi-steady state. Despite there being magnetograms extending back to 2007 in the literature, we focus only on the epoch surrounding the anticipated polarity reversal. We were specifically interested in capturing the changing space weather environment that resulted from such a drastic change in the alignment of the large-scale field. For the chosen magnetic maps, the \texttt{BATS-R-US}/\texttt{AWSoM} model permits us to simulate stellar winds out to the astrosphere, and study the space weather environment to which orbiting planets may be subject. 

In the \texttt{AWSoM} model, the inner boundary is set in the chromosphere, and the corona was heated by the dissipation of low-frequency Alfvén waves. Their propagation and partial reflection results in a turbulent cascade that heats and accelerates the solar wind~\citep{vanderHolst2014, Gombosi2018}. These Alfvén waves initially propagate outwards along the magnetic field lines. However, some are reflected due to magnetic inhomogeneities. The Alfvén waves will propagate in both parallel and antiparallel directions (relative to the magnetic field lines). It is the interaction between these counter-propagating waves that facilitates the turbulent cascade, which deposits energy into the plasma, heating it, and accelerating the stellar wind~\citep{vanderHolst2014, Gombosi2018}. The \texttt{AWSoM} model has undergone extensive validation and successfully models the solar wind
\citep{%
2014ApJ...782...81V,
2019ApJ...887...83S,
2019ApJ...872L..18V,
2023ApJ...946L..47H%
}. The model has been applied to study the winds of solar-type stars and M dwarfs~\citep[e.g.][]{2016A&A...588A..28A,2016A&A...594A..95A,2017ApJ...835..220C,2017ApJ...843L..33G,2021MNRAS.504.1511K,Evensberget2021,Evensberget2022,Evensberget2023,2022MNRAS.509.5117S,2023MNRAS.522..792M,Evensberget2024,2024MNRAS.533.1156S,Bellotti2024,2025arXiv251116370E}. For the stellar work mentioned, the wind models are driven by ZDI magnetic maps of the large-scale magnetic field. \texttt{AWSoM} shows good agreement with solar wind observations when it is driven by low resolution magnetograms that are comparable stellar ZDI magnetic maps~\citep[e.g.][]{2017ApJ...835..220C,Evensberget2021}.

\subsection{Two-temperature model equations}
In our models we make use of the two-temperature magnetohydrodynamic equations, accounting for the disparate behaviour between electrons and ions in plasma. Treating electrons and ions in this fashion has been shown to more accurately reproduce the bimodality of the wind~\citep{2017ApJ...835..220C}. The equations that comprise the two-temperature model are the conservation of mass, the conservation of momentum, the conservation of energy, and the induction equation \citep[as described in][]{vanderHolst2014}. To highlight the contribution of Alfvén wave heating to the ion and electron temperatures, we state the equations governing electron pressure ($P_\text{e}$) and ion temperature pressure ($P_\text{i}$),

\begin{align}
    \frac{\frac{\partial P_\text{i}}{\partial t}+\nabla (P_\text{i} \mathbf{u})}{\gamma - 1}+ P_\text{i} \nabla \mathbf{u} &= \frac{P_\text{e} - P_\text{i}}{\tau_{\text{eq}}}+f_\text{i}Q_\text{W} - \rho\frac{GM\mathbf{r}\cdot\mathbf{u}}{r^3}, \label{eq: Pi} \\
    \frac{\frac{\partial P_\text{e}}{\partial t}+\nabla (P_\text{e} \mathbf{u})}{\gamma - 1}+ P_\text{e} \nabla \mathbf{u} &= -\frac{P_\text{e} - P_\text{i}}{\tau_{\text{eq}}}+(1-f_\text{i})Q_\text{W}-Q_{\text{A}}-\nabla\mathbf{q}_\text{e}, \label{eq: Pe}
\end{align}
respectively. These expressions represent the conservation of energy for ions and electrons. 
Here we take $\gamma$, the ratio of specific heats, to be $5/3$. The plasma velocity is denoted by $\mathbf{u}$. On the right-hand side of Eq.~\ref{eq: Pi} and Eq.~\ref{eq: Pe}, the terms represent the collisional energy transfer, $(P_\text{i} - P_\text{e})/\tau_{\text{eq}}$, between ions and electrons; heating from Alfvén wave dissipation, $Q_\text{W} = Q_\text{W}^+ + Q_\text{W}^-$, and the fraction of ion heating, $f_\text{i}$ (we use 0.6); work against the star's gravitational potential $(\rho GM/r^3)\mathbf{r}\cdot\mathbf{u}$; and for Eq.~\ref{eq: Pe}, energy lost from an optically thin plasma as radiation, $Q_\text{A}$; and electron heat conduction, $\nabla \mathbf{q}_\text{e}$.
The total thermal pressure in a plasma is the sum of the ion pressure and the electron pressure in the normal way for the two-temperature magnetohydrodynamic equations~\citep[see e.g.][]{vanderHolst2014}.

\subsection{Alfvén wave propagation, reflection, and dissipation \label{sec: wave prop}}

Here we discuss wave energy densities and their role in heating the wind plasma. In Eq.~\ref{eq: Pe} and Eq.~\ref{eq: Pi}, we note that $Q_\text{W}^+$ and $Q_\text{W}^-$ represent the energy dissipation as an Alfvén wave propagates in parallel and antiparallel directions relative to the magnetic field, respectively. 
In the \texttt{AWSoM} model the two-temperature magnetohydrodynamic equations are extended by two further partial differential equations that describe the propagation, reflection, and dissipation of Alfvén waves energy. 
The energy densities of Alfvén waves travelling in either direction were computed from 
\begin{align}
    \frac{\partial w^\pm}{\partial t} + \nabla ((\mathbf{u}\pm \mathbf{v}_\text{A})w^\pm) + \frac{w^\pm}{2}(\nabla \cdot \mathbf{u}) = \mp R\sqrt{w^- w^+} - Q_\text{w}^\pm, \label{eq: A_wave_prop}
\end{align}
where $\mathbf{v}_\text{A}=\mathbf{B}/\sqrt{\mu_0\rho}$ is the Alfvén velocity and $\mp R\sqrt{w^- w^+}$ is the reflection rates transferring energy between the parallel and antiparallel travelling waves. This expression accounts for both the dissipation ($Q_\text{w}^{\pm}$) and reflection of the Alfvén waves. Here we can also note that the parallel and antiparallel travelling waves are related to the Alfvén wave pressure through $P_\text{A}=(w^+ + w^-)/2$. We emphasise that the \texttt{AWSoM} model numerically solves Eq.~\ref{eq: A_wave_prop} along with the two-temperature magnetohydrodynamic equations in a fully integrated manner.

\subsection{Boundary conditions \label{sec: boundary_conditions}}
The simulation domain was discretised with a spherical grid extending from the stellar chromosphere out to 250 stellar radii. Near the stellar surface, we applied geometric mesh refinement to better resolve the transition region. Additionally, in regions where the radial magnetic field, $B_\text{r}$, changes sign, we implemented automatic mesh refinement. \texttt{AWSoM} is capable of capturing the transition region, where plasma is heated to coronal temperatures, by following the approach of \citet{Oran2013} and constructing a mesh grid that is irregular in $\Delta r$ but with finer resolution in the transition region, and by numerically broadening the transition region such that it covers multiple grid cells in the radial direction~\citep{Sokolov2013}. 

For the stellar wind models of AD Leo, we took $M=0.42M_\odot$~\citep{Mann2015,Cristofari2023}, $R=0.39R_\odot$ \citep{Morin2008}, and $P_\text{rot}=2.2399$\,days~\citep{Morin2008, Fouque2018}. At the inner boundary of the model (that is, the chromosphere), we set the number density and temperature to the solar values of $2\times10^{17}$\,m$^{-3}$ and $5\times10^{4}$\,K, respectively~\citep[following][]{Gombosi2018}. The Alfvén flux-to-field ratio $(\Pi_\textsc{a}/B)$ was set to $0.77\times10^6$ Wm$^{-2}$T$^{-1}$; this value was obtained by applying the $(\Pi_\textsc{a}/B)=(\Pi_\textsc{a}/B)_\odot(R/R_\odot)^{0.3}$ scaling of~\citep{Sokolov2013} to the solar value, $(\Pi_\textsc{a}/B)_\odot=10^6$ Wm$^{-2}$T$^{-1}$, of, for example,~\citet{Gombosi2018}. The Alfvén wave transverse correlation length was set to $1.5\times10^5$\,T$^{1/2}$\,m, as in \citet{Gombosi2018}. 

As the model relaxed towards a steady state, the radial component of the boundary magnetic field was fixed to the local magnetogram value $\mathbf{B}_{\text{ZDI}}\cdot \mathbf{\hat{r}}$, i.e. the values from the first panel of Fig.~\ref{fig: map}, while the tangential magnetic field components were permitted to vary freely. 

\subsubsection{ZDI as lower bound on large-scale field \label{sec: underestimation}}

The tomographic reconstruction method applied in ZDI has been shown to under-report the strength of large-scale stellar magnetic fields~\citep{Morin2010, Lehmann2019} and is sometimes termed a lower bound on the large-scale field strength. This has led to wind modelling studies in which a scaling factor is applied to the ZDI-reported magnetic field strength~ \citep[e.g. a factor of \(1\times\) and \(5\times\) in][]{Evensberget2021,Evensberget2022,Evensberget2023} in an attempt to constrain wind properties. Meanwhile, studies aimed at uncovering relative trends typically apply unscaled ZDI maps~(e.g. \citealt{Cohen2014,2016A&A...588A..28A,2016A&A...594A..95A}, and the unscaled (\(1\times\)) series of \citealt{Evensberget2021,Evensberget2022,Evensberget2023}).
As this work is focussed on how  the changing geometry of the large-scale magnetic field (surrounding the failed-reversal) impacts the space weather environment, we use the unscaled ZDI maps.

\subsubsection{The role of the small-scale field}
The small-scale magnetic field, which is unobserved by ZDI, is considered unlikely to affect the structure of the stellar wind at planetary distances~\citep{Lang2014,2021A&ARv..29....1K}. This true in particular for polytropic magnetohydrodynamic models, in which the corona is already at megakelvin temperatures at the lower boundary~\citep[e.g.][]{2007ApJ...654L.163C,2009ApJ...703.1734V}. For the \texttt{AWSoM} model, used in this work,  it is also observed~\citep[e.g.][]{Evensberget2021,Evensberget2022,Evensberget2023} that the large-scale structure of the magnetic field is not significantly affected by small-scale magnetic flux differences. 

The \texttt{AWSoM} model is, however, somewhat more sensitive to variations in the medium- and small-scale magnetic field, as coronal heating is influenced by the magnitude of \(B\) at the inner boundary. In the \texttt{AWSoM} model, the local Alfvén wave pointing flux entering the model at a point \((\theta, \varphi)\) on the inner 
boundary is given by \(\Pi_\textsc{a}(\theta, \varphi) = (\Pi_\textsc{a}/B) \, B(\theta, \varphi)\), where \(B(\theta, \varphi)\) is the local magnetic field strength and \((\Pi_\textsc{a}/B)\) is the Alfvén flux-to-field ratio (a modelling constant; see Sect. \ref{sec: boundary_conditions}).

A direct inclusion of the small-scale field (e.g. with a flux carpet as in~\citealt{Lang2014}) would increase the local values of \(B_r\) (and thus \(B\)) beyond the values given in Fig.~\ref{fig: map}, and increase the influx of Alfvén wave energy that can be dissipated and contribute to coronal heating (see Eq.~\ref{eq: A_wave_prop}). In contrast to the flux carpet approach, there are also studies that use the mean magnetic field strength found by Zeeman broadening~\citep{2008A&A...489L..45R,2010ApJ...710..924R} to set the mean magnetic field strength of the large-scale field~\citep{Garraffo2016,2017ApJ...843L..33G}. This approach thus yields an upper bound on the large-scale field when a large-scale magnetic map is available, as it places the total magnetic energy into the large-scale field. Additionally, several studies account for variations in the small-scale magnetic flux by modifying the Alfvén flux-to-field ratio \((\Pi_\textsc{a}/B)\). 
The consequences of such a scaling have been explored by, for example, \citet{Saikia2020}, \citet{Airapetian2020}, and \citet{2023ApJ...946L..47H} and \citet{2024ApJ...965....1H}, and it has been shown to affect wind parameters such as the average Alfvén radius and the wind mass loss rate. 

Given, however, that we focus on studying the space weather effects of the changing stellar magnetic field, and to facilitate comparison with similar wind modelling studies using the \texttt{AWSoM} model~\citep{2016A&A...588A..28A,2016A&A...594A..95A,2017ApJ...835..220C,2021MNRAS.504.1511K,Evensberget2021,Evensberget2022,Evensberget2023,2021MNRAS.500.3438O,2022ApJ...924..115O,2022MNRAS.509.5117S,2023MNRAS.522..792M,Evensberget2024,2024MNRAS.533.1156S,Bellotti2024,2025arXiv251116370E}, we applied a compatible methodology and did not vary scaling factors or boundary conditions except as described in Sect. \ref{sec: boundary_conditions}. We note, however, that the role of the small-scale stellar magnetic field in stellar wind modelling remains an open field of research.

\subsection{Wind modelling results}
In this section we discuss simulation results at five epochs for the Alfvén surface structure, the radial wind velocity, the total wind pressure profiles, and stellar mass loss rates. Our primary interest is the evolution of the space weather environment over time in response to a changing magnetic field. While there are currently no confirmed detections of exoplanets orbiting AD~Leo, \citet{Vidotto2013b} place the inner and outer edges of the insulation habitable zone at 0.11\,au (56\,$R_*$) and 0.30\,au (152\,$R_*$), respectively. Throughout this section, we examine the space weather conditions at these orbital distances to understand the environment hypothetical planets may be subject to. Additionally, we examined the planetary magnetospheric stand-off distance for a planet with an Earth-like magnetic field. Given that AD~Leo is an extremely active star, it would result in more extreme space weather than slowly rotating M dwarfs with large-scale fields of the order of hundreds of Gauss~\citep{Lehmann2024}. Although we may expect quite a harsh environment around AD~Leo, what is not yet well understood is how the severity of this environment will evolve over time with the magnetic field.

\subsubsection{Radial stellar wind velocity and the Alfvén surface}
\begin{figure*}
    \centering
    \includegraphics[width=1\linewidth]{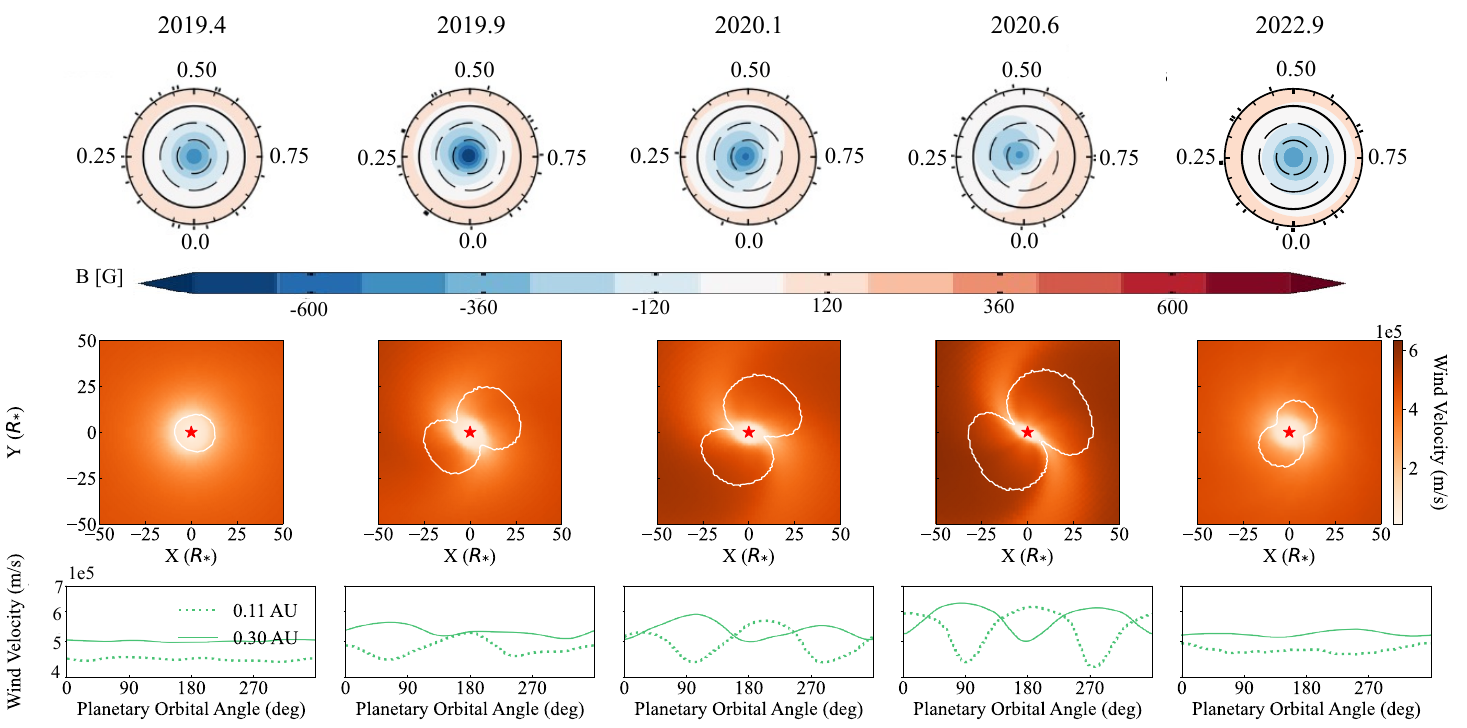}
    \caption{Top row: Magnetograms of the radial magnetic field components from (left to right) 2019.4, 2019.9, 2020.1, 2020.6, 2022.9. Middle row: Radial wind velocity in the equatorial plane, produced using the above ZDI maps. The white contour is the intersection of the Alfvén surface with the equatorial plane. We note that the scale of this plot is too small to include the orbital distance of the habitable zone. Bottom row: Radial wind velocity in the planetary frame. Wind velocity values are presented for two estimates for the habitable zone of AD~Leo: 102\,$R_*$ (dotted line) and 81\,$R_*$ (solid line).}
    \label{fig: wind_vel}
\end{figure*}

The evolution of the radial wind velocity, in the equatorial plane, around AD~Leo is displayed in the upper panels of Fig.~\ref{fig: wind_vel}. Wind speed tends to increase with distance from a star. Eventually, the wind velocity, $|u|$, will exceed what is called the Alfvén velocity - the speed with which magnetic waves can propagate through plasma. When $|u|>v_A$, disturbances from the wind cannot propagate towards the star~\citep{Alfven1947, Adhikari2019}. This transition to wind speeds greater than the Alfvén speed occurs at the Alfvén surface, $S_\text{A}$, where $u=v_\text{A}$. The interactions a planet can have with its host star are governed by whether it orbits within or beyond this surface. In the super-Alfvénic regime ($u>v_\text{A}$), stellar winds are decoupled from the magnetic field and shocks and discontinuities can arise if super-Alfvénic winds encounter obstacles such as slower winds or a planetary magnetosphere \citep[just as Earth forms a bow shock in response to solar wind;][]{Vidotto2010, Belenkaya2024}. In the sub-Alfvénic regime ($u<v_\text{A}$), where stellar winds are magnetically dominated, there can be bidirectional transport of energy between the star and planet~\citep{Strugarek2023}. Planets in such orbits can experience magnetic connectivity with the host star, which allows for the transport of energy and material between the two bodies. This connectivity can enhance stellar activity~\citep{Vidotto2023, Paul2025} and also threaten planetary atmospheric retention~\citep{Presa2024}.

\begin{figure}
    \centering
    \includegraphics[width=0.75\linewidth]{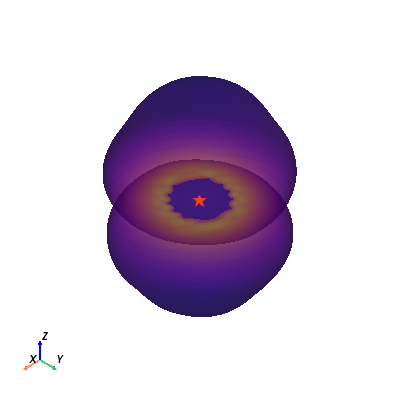}
    \caption{Three-dimensional Alfvén surface for 2022.9 shown in purple, with the position and approximate size of AD~Leo represented by the central red star.}
    \label{fig: 3D_AlfvenSurf}
\end{figure}

Superimposed on each panel of Fig.~\ref{fig: wind_vel}, in white, are two-dimensional depictions of the Alfvén surface in the equatorial plane. We observe a two-lobed structure, which is typical for stars with a dominant dipolar large-scale field topology~\citep[e.g.][]{Evensberget2021, Evensberget2022, Evensberget2023, Chebly2023, Bellotti2024}. This structure is better depicted in a three-dimensional plot (an example is shown for 2022.9 in Fig.~\ref{fig: 3D_AlfvenSurf}). We note that for 2019.4, the circular appearance of this surface is because we are only seeing a cut through the equatorial plane. When plotted in the X-Z (meridional) plane, or in 3D, where the axis of rotation is included, we unveil the two-lobed structure. The apparent size of the surface varies over time, being largest in 2020.6 before shrinking in the final epoch. These changes simply reflect variations in the axisymmetry of the large-scale field. Table~\ref{tab: wind_results} shows the average size of the Alfvén surface computed in our models. It is apparent that around AD~Leo, hypothetical habitable zone planets (at 56\,$R_*$ and 152\,$R_*$) on circular equatorial plane orbits will exist strictly in the super-Alfvénic regime. Given their considerable distance from the star, these orbits are not shown in the middle panels of Fig.~\ref{fig: wind_vel}.

The lower panels of Fig.~\ref{fig: wind_vel} show the radial wind velocity, at both of the aforementioned habitable zone distances, along the planetary orbit. We note an increase in the complexity of the wind structure, and corresponding speeds, for the wind models based on 2020 observations of AD~Leo - when the large-scale magnetic field topology exhibited a substantial reduction in axisymmetry. The observed dips in the wind speed curves correspond to each of the spiral arms shown in the middle panels. We note that the greatest complexity and variability in the wind speed structure occurs relatively close in to the star, at distances well before the habitable zone estimates. The wind speed, at all epochs, varies by less than a factor of 1.5 along the orbital distances examined - on average varying by a factor of only $\sim1.1$. At its lowest, the wind speed exhibits a value of 428447\,m/s at 56\,$R_*$ (2020.6) and a value of 494556\,m/s at 152\,$R_*$ (2019.4). Meanwhile, at its greatest, the wind speed takes on values of 613898\,m/s at 56\,$R_*$ (2020.6) and 627008\,m/s at 56\,$R_*$. The conditions are at their most variable in 2020.6, as apparent in Fig.~\ref{fig: wind_vel}, with the wind speed varying by factors of 1.4 and 1.3 for the inner and outer habitable zone distances, respectively.

\subsubsection{Total stellar wind pressure}
\begin{figure*}
    \centering
    \includegraphics[width=\textwidth]{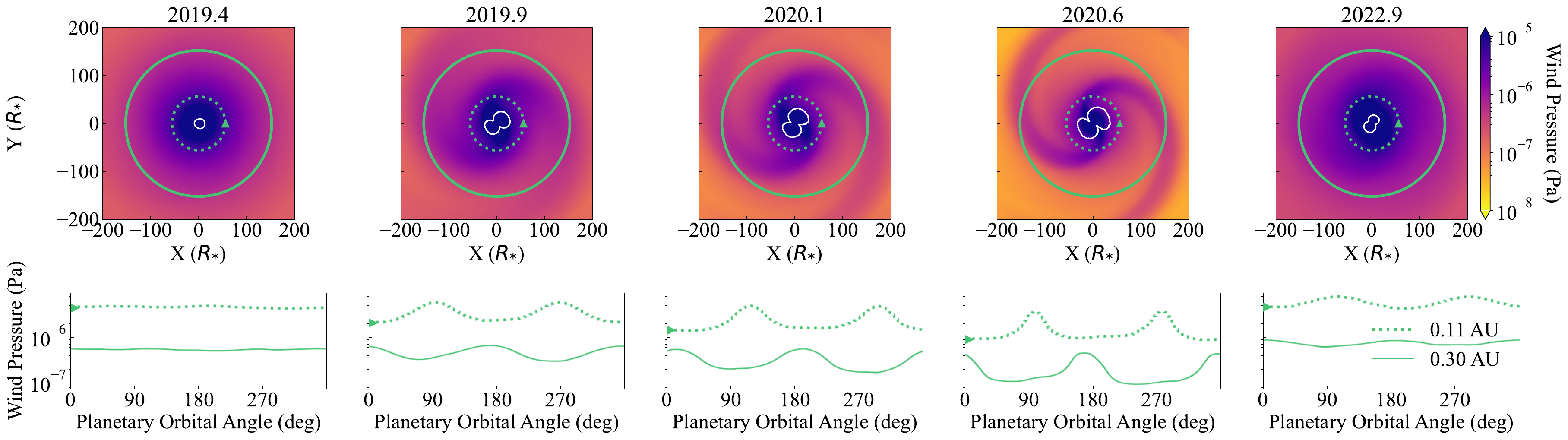}
    \caption{Top row: Total wind pressure in the equatorial plane, produced using the same ZDI maps as Fig.~\ref{fig: wind_vel}. The white contour is the intersection of the Alfvén surface with the equatorial plane. Meanwhile, the green circumferences represent the two orbital distances explored in lower panels. Bottom row: Wind pressure in the planetary frame, with the same orbits as Fig.~\ref{fig: wind_vel}.}
    \label{fig: wind_pressure}
\end{figure*}

Figure~\ref{fig: wind_pressure} shows the total wind pressure around AD Leo in the equatorial plane. As before, superimposed, in white, is the Alfvén surface. Further, habitable zone orbits are indicated as light and dark green circles. To calculate the total pressure, we followed the method of \citet{Vidotto2011}:\begin{align}
    P_W = P + \frac{|\mathbf{B}|^2}{2\mu_0} + \rho|\mathbf{u}-\mathbf{v}|^2, \label{eq: total_wind_pressure}
\end{align}
where $P=P_\text{i}+P_\text{e}$ and $\mathbf{v}$ denotes the planetary orbital velocity ($\mathbf{v}=\sqrt{GM/r}\hat{\phi}$, where $r$ is the orbital distance). As the large-scale magnetic field evolves with time, so too do the relative contributions of the thermal pressure, magnetic pressure, and ram pressure (first, second, and third term of Eq.~\ref{eq: total_wind_pressure} to the total wind pressure. At large distances, the ram pressure dominates, meanwhile closer-in the magnetic pressure dominates~\citep{Vidotto2010}. The transition point between these regimes varies depending on the epoch, though it occurs at an average distance of $9$\,$R_*$. In every model, the habitable zone orbits are placed at distances where ram pressure dominates, just as for the Solar System planets.

Three out of the five models (2019.9, 2020.1, and 2020.6) produced in this work show a two-armed spiral structure typical of a tilted large-scale dipolar field \citep{Evensberget2023}. Much like our results for the wind velocity, we observe greater complexity in the observed structures for the aforementioned models. Tracing the values along the hypothetical planetary orbits (perpendicular to the stellar rotation axis) at the habitable zone boundaries, we can better see the influence of the complexity of the wind pressure structures on the environment to which a potential planet would be subject. For the 2019.4 and 2022.9 models, potential planets on either orbit would experience less variable conditions, yet almost consistently greater wind pressure values. In every case, the wind pressure varies by a greater amount (a factor of $\sim2.6$ on average) along a planet's orbit than the wind speed. At its lowest, the wind pressure has values of $\sim8.6\times10^{-7}$\,Pa at 56\,$R_*$ and $\sim9.2\times10^{-8}$\,Pa at 152\,$R_*$. Meanwhile, at its greatest it has values of $\sim3.6\times10^{-6}$\,Pa and $\sim4.5\times10^{-7}$\,Pa. Notably, each of these extremes occurs in 2020.6, where, once again, we observe the greatest equatorial plane variability. At this epoch, wind pressure values vary by approximately factors of 4.1 and 4.8 at 56\,$R_*$ and 152\,$R_*$, respectively. 

At the distances examined the total wind pressure is ram dominated -- calculated as the interaction between wind speed and wind density provided by the last term of Eq.~\ref{eq: total_wind_pressure}. For a hypothetical planet located at these distances, the ram dominated nature of the total stellar wind pressure would exert compressive forces on planetary magnetospheres, which may reduce their size. For unmagnetised, or weakly magnetised planets, these pressures might drive atmospheric erosion through ion pick-up and sputtering processes~\citep{Kislyakova2014, Leblanc2018}. 

\subsubsection{Planetary magnetospheric stand-off distance}
The direct interaction between stellar winds and planetary atmospheres may result in atmospheric erosion. However, atmospheres can be sheltered if a planet possesses a large-scale magnetic field, which provides sufficient magnetic pressure to prevent direct interactions between stellar winds and the planetary atmosphere~\citep{Chapman1930, Veras2021}. In super-Alfvénic regimes, a planet's magnetosphere produces a bow shock that deflects the stellar winds that confront it. We can quantify the distance between a planet and the dayside shock with the magnetospheric stand-off distance. This quantity provides us with an indication of the likelihood of atmospheric erosion, although this idea has recently been challenged in the literature (see below). 

The planetary magnetospheric stand-off distance, $R_\text{m}$, can be calculated using the expression derived from a pressure balance argument~\citep[e.g.][]{Vidotto2009}:
\begin{align}
    \frac{R_\text{m}}{R_\text{p}} = \left(\frac{P_\text{p}}{P_{\text{ram}}}\right)^\frac{1}{6} = \left(\frac{B_0^2/(2\mu_0)}{\rho u^2}\right)^\frac{1}{6}.
\end{align}
Here, $P_\text{p} = B_0^2/(2\mu_0)$ is the planetary magnetic pressure at the surface, with $B_0$ being the strength of the planetary field (which, for an Earth-like planet, is 0.7\,G), and $R_\text{p}$ is the planetary radius. \citet{Lammer2007} found that, to prevent atmospheric erosion, the magnetospheric stand-off distance for a planet orbiting an M dwarf should be greater than 2 planetary radii. For AD~Leo, we find that the smallest value obtained, in either habitable zone distance at any epoch, is 2.7\,$R_\text{p}$ (Table~\ref{tab: wind_results}). As we discuss in Sect. \ref{sec: sum}, we do not obtain any sub-2\,$R_\text{p}$ values until approximately a planetary field strength of 0.34\,G. Similarly to the previous quantities examined, we find greater variability in $R_\text{m}$ for 2019.9, 2020.1, and 2020.6. However, at all epochs the variation along a planet's orbit is less than a factor of 1.5. The greatest variability occurs for 2020.6 where, at both distances, $R_\text{m}$ values vary by approximately a factor of 1.3. Thus, if the metric suggested in \citet{Lammer2007} can indeed be a good indication for the prevention of substantial atmospheric erosion, Earth-like planets orbiting in the habitable zone of AD~Leo would not (presently) have their atmospheres eroded by the stellar wind. It is worth noting, though, that there is currently a debate in the literature questioning whether planetary magnetic fields can indeed protect atmospheres of terrestrial planets \citep[see, for example,][]{Gunell2018, Blackman2018, Egan2019, Sakata2022}.

\subsubsection{Stellar wind mass loss rates}
The stellar wind mass loss rate is the integral of the mass flux over a closed surface centred on the star: 
\begin{align}
    \dot{M} = \oint_S \rho\mathbf{u}\cdot \mathbf{n}~\mathrm{d}S \label{eq: mass loss rate}
,\end{align}
where $\mathbf{n}$ is the unit vector normal to the surface, $S$. If we have a steady state solution, the value of $\dot{M}$ should be independent of the surface, $S$, granted the surface encloses the star (for quasi-steady state solutions this independence is not guaranteed).

\begin{figure}
    \centering
    \includegraphics[width=1\linewidth]{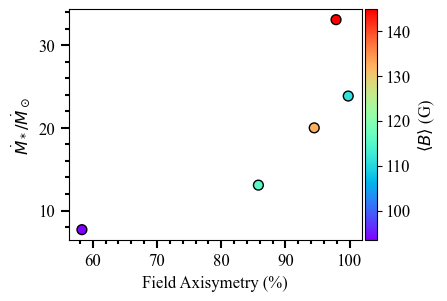}
    \caption{Stellar mass loss rate \citep[as a fraction of the solar mass loss rate of $1.3\times10^9$\,kg/s,][]{Mishra2019} against the large-scale magnetic field axisymmetry of AD~Leo, at epochs 2019.4, 2019.9, 2020.1, 2020.6, and 2022.9. The points are coloured by the mean large-scale magnetic field strength obtained from ZDI large-scale field reconstructions at each epoch.}
    \label{fig: massloss}
\end{figure}

We find stellar mass loss rates of the order of $10^{10}$\,kg\,s$^{-1}$, with the greatest value being $4.3\times10^{10}$\,kg\,s$^{-1}$ in 2022.9 and the smallest value being $1.0\times10^{10}$\,kg\,s$^{-1}$ in 2020.6 (see Table~\ref{tab: wind_results}). These values are an order of magnitude larger than the solar mass loss rate, which has been estimated to be, on average, $1.3\times10^9$\,kg/s~\citep{Mishra2019}. We see no clear correlation between the stellar mass loss rates and stellar large-scale magnetic field strength evolution. Although there are too few large-scale magnetic field strength values to truly examine this relationship. Additionally, any correlations between stellar mass loss rates and large-scale field complexity are difficult to discern. The large-scale field of AD~Leo, at all epochs, is predominantly dipolar and so such a relationship is challenging to assess. However, we do find that lower values of large-scale field axisymmetry tend to correspond to lower mass loss rates, Fig.~\ref{fig: massloss}.\\ 

\begin{table}[!t]
\caption{Overview of wind quantities examined.}
\label{tab: wind_results}
\centering
\small 
\setlength{\tabcolsep}{4pt}
\begin{tabular}{l l c c c c c}
\hline
Quantity & Units & 2019.4 & 2019.9 & 2020.1 & 2020.6 & 2022.9 \\
\hline
$R_A$ & $R_*$ & 7.3 & 16.1 & 18.5 & 21.3 & 11.0 \\
$\dot{M}$ & $10^{10}$\,kg\,s$^{-1}$ & 3.1 & 2.6 & 1.7 & 1.0 & 4.3 \\
$\langle P_w^{56R_*} \rangle$ & $10^{-6}$\,Pa & 4.5 & 3.3 & 2.3 & 1.4 & 5.7 \\
$\langle P_w^{152R_*} \rangle$ & $10^{-7}$\,Pa & 5.3 & 4.5 & 3.1 & 1.9 & 7.4 \\
$\langle R_m^{56R_*} \rangle$ & $R_p$ & 2.8 & 3.0 & 3.2 & 3.5 & 2.7 \\
$\langle R_m^{152R_*} \rangle$ & $R_p$ & 3.9 & 4.1 & 4.4 & 4.8 & 3.7 \\
\hline
\end{tabular}
\tablefoot{$R_A$ denotes the distance to the Alfvén surface averaged over the stellar surface, $\dot{M}$ is the stellar mass loss rate, $P_w^{56R_*}$ and $P_w^{152R_*}$ are the wind pressure averaged over hypothetical planetary orbits at 56\,$R_*$ and 152\,$R_*$, and $R_m^{56R_*}$ and $R_m^{152R_*}$ are the planetary magnetospheric stand-off distances averaged over hypothetical planetary orbits at 56\,$R_*$ and 152\,$R_*$.}
\end{table}

\begin{figure}
    \centering
    \includegraphics[width=1\linewidth]{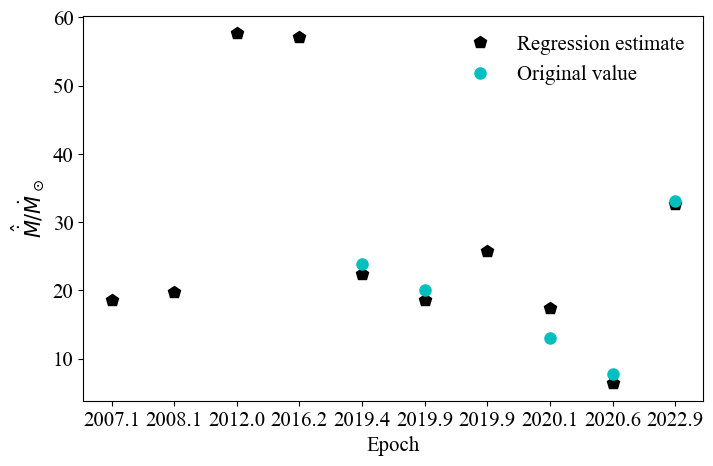}
    \caption{Stellar mass loss rate \citep[as a fraction of the solar mass loss rate of $1.3\times10^9$\,kg/s,][]{Mishra2019} of AD~Leo at across epochs spanning from 2007 to 2022. The black points represent mass loss rate estimates obtained from Eq.~\ref{eq: regression} and the blue points display the original values from Table~\ref{tab: wind_results}.}
    \label{fig: Mdot_regression_est}
\end{figure}

Although we have not simulated the winds for the optical data, we can provide a rough estimate of its values based on the simulation results for the NIR data. To do this, we performed a multivariable regression - focussing on the impacts of $\langle B \rangle$, $\text{f}_\text{dip}$, and $\text{f}_\text{axisym}$ (from Table~\ref{tab: zdi_output}) on the mass loss rate.
From doing so, we arrive at the equation
\begin{align}
    \hat{\dot{M}} = a \langle B\rangle+b\text{f}_\text{dip} +c \text{f}_\text{axisym} + d, \label{eq: regression}
\end{align}
where $a$, $b$, and $c$ denote the coefficients for $\langle B \rangle$, $\text{f}_\text{dip}$, and $\text{f}_\text{axisym}$, respectively, and $d$ is the regression intercept. We find values of $a=(1.89\pm33.16)\times10^8\,\text{kg s}^{-1}\text{G}^{-1}$, $b=(7.66\pm64.25)\times10^8\,\text{kg s}^{-1}$, $c=(2.16\pm38.13)\times10^8\,\text{kg s}^{-1}$, and $d=(-7.56\pm37.77)\times10^{10}\,\text{kg s}^{-1}$. The results from the regression, applied to the optical and NIR data, are shown in Fig.~\ref{fig: Mdot_regression_est}.

Ultimately, applying Eq.~\ref{eq: regression} to the data from the NIR ZDI maps returns comparable values to the mass loss rate estimates obtained from the wind models and Eq.~\ref{eq: mass loss rate}. However, we do note some minor discrepancies, which likely arise due to the relatively unclear trend depicted by the five data points used in the regression. Due to the unclear trend, and limited number of data points used in the regression, we exercise caution in our analysis of the extrapolated and interpolated optical values. Our intention here is only to provide rough estimates.

We find that the mass loss rates in 2007.1 and 2008.1 resemble those calculated for 2019.4 and 2019.9 (NIR). These epochs differed across the three quantities explored (i.e. $\langle B \rangle$, $\text{f}_\text{dip}$, and $\text{f}_\text{axisym}$), highlighting the complex interplay between them that governs the stellar wind mass loss rate. Epochs 2012.0 and 2016.2 displayed mass loss rates almost double that of the next highest measurement (2022.9). These epochs were characterised by comparable $\text{f}_\text{dip}$ and $\text{f}_\text{axisym}$ values to 2022.9; however, they possessed almost double the average large-scale field strength. Across all the epochs, we find the mass loss rate reaches its lowest value in 2020.6, as the large-scale magnetic field of AD~Leo borders on a polarity reversal.

\subsection{Scaling the magnetic field \label{sec: scaling}}
AD~Leo is an active star and, as such, possesses a stronger magnetic field than inactive M dwarfs. In this section we consider how our results might change if the large-scale field strength values at each epoch were to be scaled down, and thus resemble those of slowly rotating M dwarfs \citep{Lehmann2024}. \citet{Evensberget2023} simulate the winds for a range of solar-like stars - comparing the results produced from unscaled ZDI maps to those produced from ZDI maps where the field strength has been scaled up by a factor of 5. As a result of this scaling: the Alfvén radius was found to increase, on average, by a factor of 1.6; the stellar wind pressure increased by a factor of $\sim$2.1-4.9 (with an average of 3.8); the planetary magnetospheric stand-off distance faced a reduction of 20\% on average; and the stellar wind mass loss rate increased by a factor of $\sim$2.2-4.3 (with an average of 3.1).

Although AD~Leo is not a solar-like star, we expect similar results from scaling. Thus, if we were to instead scale the field strength down by a factor of 5, we would expect the reductions in the Alfvén radius, stellar wind pressure, and planetary magnetospheric stand-off distance, and an increase in the stellar wind mass loss rate, by the aforementioned factors. Ultimately, we would arrive at values on the same order of magnitude as our current estimates.

\section{Summary and discussion \label{sec: sum}}
In this work, we follow up on the anticipated polarity reversal of AD Leo. With new spectropolarimetric observations, we used ZDI to reconstruct the photospheric large-scale magnetic field. The work of \citet{Bellotti2023b} noted a significant reduction in the axisymmetry of the large-scale field in 2020, hinting at the potential for a reversal. However, our new magnetic map suggests a highly dipolar and axisymmetric configuration. This failed reversal is not an isolated incident, however. Rather, dipole excursions \citep[i.e. the temporary weakening and tilting of a dipolar large-scale field;][]{Gubbins1999, Valet2005} have been studied extensively (both experimentally and with numerical simulations) in the context of geomagnetism and there is now growing interest in their implications in stellar contexts~\citep{Wicht2004, Petrelis2009, DeRosa2012, Augustson2013}. \citet{Bellotti2025} identify a G2 star, HD~76151, which exhibited oscillatory variations in axisymmetry for more than a decade in the lead-up to a singular polarity reversal. Further, the work of \citet{Brown2024} highlights the failed polarity reversal of another G2 star, V899~Her. And so, future spectropolarimetric observations of AD~Leo will be required to determine whether it will eventually undergo a complete polarity reversal.

Using four magnetic maps produced in \citet{Bellotti2023b}, and the map produced in this work, we then simulated the space weather environment around the star over time using the Space Weather Modelling Framework (\texttt{SWMF}). We evaluated simulated results for the radial wind velocity and the total wind pressure in the equatorial plane, as well as the location of the Alfvén surface, the planetary magnetospheric stand-off distance, and stellar mass loss rates. 

Due to the dipolar-dominated large-scale fields of AD~Leo, the Alfvén surface at each epoch is characterised by a two-lobed structure, as expected~\citep{Evensberget2021,Evensberget2022,Evensberget2023}. However, we note an evolution in the size of the Alfvén surface as the large-scale field evolves, with the surface fluctuating in size. We found that hypothetical planets orbiting on circular trajectories in the equatorial plane, at habitable zone distances, would orbit strictly in the super-Alfvénic regime. In this regime, a planetary magnetosphere would produce a bow shock, when confronted by stellar winds, which would deflect stellar winds and could help preserve a planetary atmosphere~\citep{Chapman1930, Vidotto2009}. Furthermore, we would not observe any planet-induced radio emission~\citep{Kavanagh2021} or local chromospheric flux enhancement~\citep{Shkolnik2008, Lanza2009}, characteristic of sub-Alfvénic star-planet interactions.

In most epochs, we can identify two spiral arms in the structure of the wind speed and pressure. These spiral arms appear to correspond to epochs with reduced large-scale field axisymmetry. Thus, as the magnetic poles tilt further from the rotation axis the space weather environment becomes increasingly variable. For more complex large-scale fields, we might expect to see additional spiral arms~\citep{Evensberget2021}. Strong stellar winds may mitigate atmosphere escape by substantially reducing erosion rates~\citep{Christie2016,Carolan2020, Vidotto2020}. However, this effect is more significant for stellar mass loss rates several orders of magnitude greater than that of our Sun, which we have not found to be the case for AD~Leo. This conclusion is also more applicable to much more close-in planets. Across the epochs, we note changes in the dominating term of the total wind pressure. The habitable zone estimates for AD~Leo are far enough from the star that, at all epochs, the ram pressure is the dominant contribution to the total stellar wind pressure.

As the wind speed and pressure change, so too does the magnetospheric stand-off distance for a magnetised planet. A large magnetosphere could shield planetary atmospheres from stellar winds and high-energy space weather events. However, the pressure exerted by stellar winds can compress the magnetosphere. As a planet's magnetosphere compresses, the atmosphere may become exposed to the space weather environment. To prevent atmospheric erosion, it is expected that a $R_\text{m}$ distance greater than 2 planetary radii is required~\citep{Lammer2007}. For a planet with an Earth-like magnetic field of 0.7\,G, we found the stand-off distances range, on average, from $2.7-4.5$\,$R_\text{p}$ at 56\,$R_*$ and from $3.7-4.8$\,$R_\text{p}$ at 152\,$R_*$. However, it is understood that the magnetosphere size (and thus the stand-off distance) increases with stronger planetary magnetic fields~\citep{Chapman1930, Cohen2020}. And so, planets with weaker magnetic fields may find themselves with stand-off distances too small to protect their atmosphere. At a planetary field strength of approximately 0.34\,G, a planet at 56\,$R_*$ in 2022.9 would have a sub-2\,$R_\text{p}$ magnetospheric stand-off distance for $\sim12\%$ of its orbit. For this planetary field strength, however, all other epochs (at both orbital distances) produce $R_\text{m}$ values greater than 2\,$R_\text{p}$. At 0.20\,G there is a drastic change in the magnetospheric stand-off distances. For this planetary field strength, all epochs at 56\,$R_*$ will have sub-2\,$R_\text{p}$ magnetospheric stand-off distances for some portion of the orbit. In the case of 2019.4 and 2022.9, we strictly observe $R_\text{m}<2$\,$R_\text{p}$. Meanwhile, for 2019.9, 2020.1, and 2020.6, we have $R_\text{m}<2$\,$R_\text{p}$ for approximately 54\%, 28\%, and 12\% of the orbits, respectively. Meanwhile, at 152\,$R_*$ all epochs, once again, strictly satisfy $R_\text{m}>2$\,$R_\text{p}$. It is not until a planetary field strength of approximately 0.11\,G that orbits at 152\,$R_*$ begin to exhibit sub-2\,$R_\text{p}$ magnetospheric stand-off distances (i.e. 43\% for 2022.9). With this, for weakly magnetised planets we can expect more severe atmospheric erosion -- which may jeopardise habitability -- at the inner boundary of the habitable zone than at the outer boundary. However, for Earth-like magnetic field strength, planetary atmospheres may be sufficiently sheltered from AD~Leo's quiescent winds across the entire habitable zone.

It has been shown that stronger stellar magnetic fields drive greater stellar mass loss rates~\citep{Cohen2014} - regardless of spectral type~\citep{Chebly2023}. The large-scale field strength of AD~Leo has not varied significantly enough, between 2019 and 2022, to critically evaluate this claim, though there does not appear to be a strict relationship between stellar large-scale magnetic field strength and stellar mass loss rate. Our observations do, however, corroborate with the work of \citet{Vidotto2013}, who found that non-axisymmetric large-scale fields produce an asymmetric mass flux. Using a multivariable regression, we were able to provide estimates for the mass loss rates for the epochs studied in the optical domain (for which we did not compute the wind modelling simulations). The regression results suggest the mass loss rate is dictated by a complex interplay between the average large-scale field strength, the contribution of the dipolar field, and the large-scale field axisymmetry.

Finally, our results were obtained for a specific M dwarf. Given the recent evidence that M dwarfs exhibit long-term variations, it may be interesting to study any correlated variation in the space weather environment around them. This will allow us to better characterise the temporal variations of quantities such as wind velocity, pressure, and mass loss rates across a wider stellar parameter space.

\section*{Data availability}

The spectropolarimetric observations analysed in this work are available on online databases. The ESPaDOnS and SPIRou observations are publicly available on PolarBase\footnote{\url{https://www.polarbase.ovgso.fr/}} \citep{Petit2014}. The respective IDs of the runs are 22BD55 and 22BD09.

\begin{acknowledgements}
    We thank the Leiden/ESA Astrophysics Program for Summer Students for hosting KGS while working on this project. 
    DE and AAV acknowledge funding from the European Research Council (ERC) under the European Union’s Horizon 2020 research and innovation programme (grant agreement No 817540, ASTROFLOW). AAV acknowledges funding from the Dutch Research Council (NWO), with project number VI.C.232.041 of the Talent Programme Vici.
    SB and AAV acknowledge funding by the Dutch Research Council (NWO) under the project "Exo-space weather and contemporaneous signatures of star-planet interactions" (with project number OCENW.M.22.215 of the research programme "Open Competition Domain Science- M").
    This work used observations obtained at the Canada-France-Hawaii Telescope (CFHT) which is operated by the National Research Council (NRC) of Canada, the Institut National des Sciences de l’Univers of the Centre National de la Recherche Scientifique (CNRS) of France, and the University of Hawaii. The observations at the CFHT were performed with care and respect from the summit of Maunakea which is a significant cultural and historic site. We gratefully acknowledge the CFHT QSO observers who made this project possible. 
    This research has made use of NASA's \href{https://ui.adsabs.harvard.edu/}{Astrophysics Data System}.
    This work used the Dutch national e-infrastructure with the support of the SURF Cooperative using grant nos. EINF-7488. 
    This work was carried out using the SWMF and BATS-R-US tools developed at the University of Michigan’s \href{https://clasp.engin.umich.edu/research/theory-computational-methods/center-for-space-environment-modeling/}{Center for Space Environment Modeling (CSEM)}. The modeling tools are available through the University of Michigan for download under a user license; an open source version is available at
    \url{https://github.com/SWMFsoftware}.
    This work has made use of the following additional software:
    Astropy 12 a community-developed core Python package for Astronomy~\citep{AstropyCollaboration2013, AstropyCollaboration2018}; NumPy~\citep{vanderWalt2011}; Matplotlib: Visualization with Python~\citep{Hunter2007}; SciPy~\citep{Virtanen2020}.
\end{acknowledgements}

\bibliographystyle{aa}
\bibliography{aa55330-25corr}

\appendix

\onecolumn
\section{Observing log}
\begin{longtable}{lccrrcccc}
\caption{\label{tab: obs} AD~Leo observations using ESPaDOnS and SPIRou. } \\
\hline\hline
Date & UT & HJD & $B_\ell$ & $n_\mathrm{cyc}$ & $t_{exp}$ & S/N & $\sigma_\mathrm{LSD}$ & Instrument\\
   & [hh:mm:ss] & [$-2450000$] & [G] &   & [s] &    & [$10^{-5}I_c$] & \\
\hline
\endfirsthead
\caption{Continued.}\\
\hline\hline

\hline
\endhead
\hline
\endfoot
\multicolumn{9}{c}{2022}\\
\hline
October 16   &   14:43:27  & 9869.1106 & $-157.5\pm8.5$  & $-11.23$ &    4x160  & 194 & 13.97  & ESPaDOnS\\
October 17   &   15:41:15   & 9870.1508 & $-104.9\pm9.0$ & $-10.32$ &    4x160  & 184 & 14.86  & ESPaDOnS\\
October 18   &   15:20:30  & 9871.1361 & $-139.7\pm8.3$ &  $-10.76$ &   4x160  & 193 & 16.06 & ESPaDOnS\\
October 19   &   15:25:33  & 9872.1400 & $-134.3\pm11.4$ &  $-9.87$ &    4x160  & 155 & 19.62  & ESPaDOnS\\
November 10       &    15:42:33  &  9894.1534 & $-157.4\pm9.6$ & 0.00 &     4x61     & 415 & 8.92 & SPIRou \\
November 11       &    15:37:26  & 9895.1500  & $-148.1\pm9.8$ & 0.45 &     4x61     & 396 & 8.20 & SPIRou \\
November 12 & 15:49:30 & 9896.1584 & $-121.6\pm9.4$ & 0.90 & 4x61 & 477 & 9.78 & SPIRou \\
November 13 & 15:34:01 & 9897.1478 & $-105.5\pm9.5$& 1.34 & 4x61 & 408 & 7.35 & SPIRou \\
November 15 & 15:31:52 & 9899.1465 & $-109.9\pm9.8$ & 2.24 & 4x50 & 397 & 8.71 & SPIRou \\
November 16 & 15:38:17 & 9900.1510 & $-131.2\pm11.8$ & 2.69 & 4x55 & 403 & 8.71 & SPIRou \\
November 17 & 15:37:24 & 9901.1505 & $-115.1\pm14.7$ & 3.14 & 4x61 & 340 & 10.47 & SPIRou \\
November 18 & 15:44:38 & 9902.1556 & $-138.4\pm9.2$ & 3.59 & 4x61 & 322 & 9.08 & SPIRou \\
November 19 & 15:35:55 & 9903.1497 & $-164.6\pm10.1$ & 4.03 & 4x61 & 349 & 10.51 & SPIRou \\
November 21 & 15:26:32 & 9905.1434 & $-131.7\pm11.4$ & 4.93 & 4x61 & 396 & 8.83 & SPIRou \\
November 22 & 14:45:53 & 9906.1152 & $-115.3\pm9.3$ & 5.36 & 4x50 & 399 & 8.00 & SPIRou \\
\hline
\multicolumn{9}{c}{2023}\\
\hline
January 6     &    15:37:03    & 9951.1548 & $-115.8\pm9.6$ &  25.56 &    4x61     & 261 & 13.76 & SPIRou \\
January 6     &    16:09:08    & 9951.1771 & $-111.5\pm9.1$ &  25.57 &     4x61     & 430 & 8.16 & SPIRou \\
January 8     &    15:14:43   & 9953.1394 & $-148.6\pm11.5$ &   26.45 &     4x61     & 379 & 8.98 & SPIRou \\
January 9    &    15:33:14   & 9954.1523 & $-100.4\pm10.7$ &   26.91 &     4x61     & 373 & 9.08 & SPIRou \\
January 10   &    15:44:13   & 9955.1557 & $-129.7\pm9.6$ &   27.36 &     4x55    & 395 & 8.31 & SPIRou \\
January 11    &    15:39:13  & 9956.1566 & $-94.9\pm9.9$ &    27.80 &     4x61     & 431 & 7.68 & SPIRou \\
January 12    &    15:13:14   & 9957.1386 & $-159.6\pm9.7$ &   28.24 &     4x61     & 389 & 8.51 & SPIRou \\
January 13    &    14:51:03   & 9958.1233 & $-135.4\pm10.0$ &   28.69 &     4x61     & 419 & 7.42 & SPIRou \\
\hline
\end{longtable}
\tablefoot{The columns are as follows: (1 and 2) date and universal time of the observations, (3) heliocentric Julian date of the observation, (4) longitudinal magnetic field strength, (5) rotational cycle of the observations calculated using Eq.~\ref{eq: ephemeris}, (6) exposure time of a polarimetric sequence, (7) S/N at 1650 nm per 2.3 km/s velocity bin and per polarimetric sequence for SPIRou and at 647 nm per 2.6 km/s velocity bin and per polarimetric sequence for ESPaDOnS, (8) RMS noise level of Stokes V signal (per 1.8 km/s velocity bin) in units of unpolarised continuum, and (9) the instrument used.}

\section{Phase-folded longitudinal field strength}
\begin{figure}[H]
    \centering
    \includegraphics[width=0.5\linewidth]{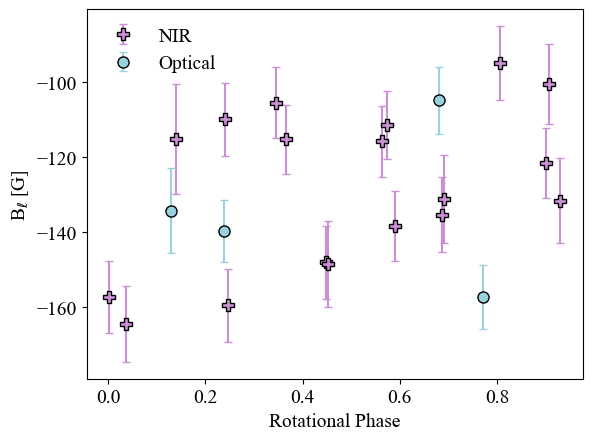}
    \caption{Phase-folded longitudinal magnetic field ($B_\ell$) measurements of AD~Leo for the 2022.9 observations. The blue circles represent observations made in the optical domain, meanwhile the pink pluses represent near-infrared observations.}
    \label{fig: phasefolded}
\end{figure}

\end{document}